\documentstyle[aps]{revtex}

\begin{document}
\draft




\title{Charge ordering in doped manganese oxides: 
 lattice dynamics and magnetic structure}

\author{J. D. Lee$^a$ and B. I. Min$^{a,b}$ }
\address{ $^a$Department of Physics,
          Pohang University of Science and Technology,
          Pohang 790-784, Korea\\
          $^b$Max-Planck-Institut f\"{u}r Festk\"{o}rperforschung,
                D-70506 Stuttgart, Germany }

\maketitle

\begin{abstract}

Based on the Hamiltonian of small polarons with the 
strong nearest neighbor repulsion, we have investigated the 
charge ordering phenomena observed in 
half-doped manganites R$_{1/2}$A$_{1/2}$MnO$_3$.
We have explored possible consequences of the charge ordering 
phase in the half-doped manganites.
First, we have studied the renormalization of the sound velocity
around $T_{CO}$, considering the acoustic phonons coupled to 
the electrons participating in the charge ordering.
Second, we have found a new antiferromagnetic phase induced
by the charge ordering, and discussed its role in connection with 
the specific CE-type antiferromagnetic structure observed 
in half-doped manganites. 

\end{abstract}



%
%


Since the discovery of colossal magnetoresistance (CMR) 
compounds R$_{\rm 1-x}$A$_{\rm x}$MnO$_3$
(R = La, Pr, Nd; A = Ca, Ba, Sr, Pb)\cite{Jin}, 
there have been extensive experimental and theoretical efforts
to find the origin of the anomalous magnetotransport phenomena.
Despite such intense studies, however, a complete understanding of 
unusual physical properties in perovskite manganites
is still lacking. 
Undoped RMnO$_3$ is an antiferromagnetic (AFM) insulator,
but with carrier doping of divalent A elements in R sites,
the system becomes a ferromagnetic (FM) metal with the observed CMR 
for 0.2 $<$ x $<$ 0.5.
Qualitative explanation is given by the double exchange (DE) 
mechanism \cite{Zener}.
At higher doping (x $\gtrsim$ 0.5), the ground 
state becomes again an AFM insulator\cite{Wollan,Schiffer}.
A phase boundary between the FM metallic and the AFM insulating 
ground states exists in a narrow range around x = 0.5\cite{Ramirez}.
In addition, another intriguing phase, the charge ordered (CO) state 
has been found to exist in insulating R$_{1/2}$A$_{1/2}$MnO$_3$ 
(RA=LaCa\cite{Radaelli}, PrSr\cite{Knizek}, PrCa\cite{Jirak},
NdSr\cite{Kuwahara}).
A direct evidence of the CO state is provided by the 
electron diffraction for La$_{1/2}$Ca$_{1/2}$MnO$_3$\cite{Chen}.

The CO state is characterized by the real-space 
ordering of Mn$^{3+}$/Mn$^{4+}$ in mixed valent R$_{1/2}$A$_{1/2}$MnO$_3$.
The CO state is expected to become stable when the 
repulsive Coulomb interaction between carriers dominates over
the kinetic energy of carriers\cite{Tomioka}.
In this respect, the electron lattice of the CO state
may be viewed as the generalized Wigner crystal.
Furthermore, the carriers to form the CO state
are believed to be manifested in some type of polarons,
which arise from the strong electron-phonon
interaction, possibly, the Jahn-Teller effect\cite{Zhao}. 
In fact, ordering of such polarons are occasionally observed in $3d$ transition 
metal oxides. The Verwey transition in Fe$_3$O$_4$\cite{Verwey} is
a typical example of the real-space ordering of mixed valent
Fe$^{2+}$ and Fe$^{3+}$ species. 
Also the formation of polaron lattice has been reported in
La$_{\rm 2-x}$Sr$_{\rm x}$NiO$_4$ (x = 1/3 and 1/2)\cite{Chen2}.

Recently, Ramirez {\it et al.}\cite{Ramirez} 
observed in La$_{\rm 1-x}$Ca$_{\rm x}$MnO$_3$ 
($0.63\leq{\rm x}\leq0.67$) that the CO transition is accompanied by 
a dramatic increase ($\gtrsim$10\%) in the sound velocity, 
implying a strong electron-phonon coupling.
Another interesting aspect of the CO state is its relevance to
the observed magnetic phases.
In the half-doped LaCa\cite{Radaelli,Chen} and PrSr\cite{Knizek} manganites,
the CO state has been realized with FM-AFM transition. 
On the other hand, in PrCa\cite{Jirak} manganites, it has been found 
in the paramagnetic (PM) state followed by the AFM transition thereafter.
However, the common feature is that the AFM structure of the specific
CE-type\cite{Wollan} is observed in the CO state of manganites,
suggesting a nontrivial effect of the CO state on the magnetic
phase.  The other noteworthy observation is the transport phenomena
of the CO phase in the presence of the magnetic field.
The high magnetic field induces the melting of the electron 
lattice of the CO phase to give rise to a huge negative 
magnetoresistance (MR) \cite{Tokura}.

In this paper, we have studied the CO transition in half-doped manganites
in the context of an order-disorder transition in interacting small polarons.
The model seems quite plausible, in view of the fact that
the transport behaviors in half-doped CMR compounds are qualitatively 
similar to the small polaron conduction in Fe$_3$O$_4$\cite{Ihle}.
We have then investigated interesting physical consequences associated with
the CO state. First, we have studied the CO effects on the 
lattice dynamics, taking into account the acoustic phonons coupled to
the CO states.  Second, considering the spin interaction in terms of
the DE, we have studied the magnetic structure resulting from
the CO state. It is found that the CO transition produces a large
sound velocity renormalization around $T_{CO}$, and that the CO state
favors the specific AFM structure apparently very similar to 
the CE-type AFM structure of half-doped manganites.

%
%


To describe the charge ordering phenomena 
in the mixed valence systems with Mn$^{3+}$:Mn$^{4+}$=1:1, 
we start with the following Hamiltonian 
of repulsively interacting small polarons,
\begin{equation}
{\cal H}_c=V\sum_{<ij>}c_i^{\dagger}c_ic_j^{\dagger}c_j
        +t\langle {\rm cos}\frac{\theta}{2}\rangle
         \sum_{<ij>}c_i^{\dagger}c_j
\end{equation}
where $<ij>$ denotes the nearest neighbor pair. Here, $V$ and $t$ represent, 
respectively, the nearest neighbor polaron interaction and the
hopping parameter, both of which are renormalized by the strong 
electron-phonon interaction\cite{Alex}.
As the carrier concentration increases, the interaction between carriers
will play a more decisive role in the system.
The second term in ${\cal H}_c$ is the DE interaction with
the single band, and the thermodynamic average
$\langle {\rm cos}\frac{\theta}{2}\rangle$ is determined
by underlying spin structures. 
We assume that electrons are strongly trapped
into small polarons so that there is no significant
overlap of orbitals. Then the DE term can be treated as
a small perturbation ($V\gg t\langle {\rm cos}\frac{\theta}{2}\rangle$).
This assumption is consistent in that the ordering of mixed valent
Mn$^{3+}$/Mn$^{4+}$ atoms weakens the DE interaction,
as the degeneracy in the ground state (Mn$^{3+}$$-$O$^{2-}$$-$Mn$^{4+}$
and Mn$^{4+}$$-$O$^{2-}$$-$Mn$^{3+}$) is lifted \cite{Goodenough}.
We also consider spinless electrons, since the strong on-site 
interaction $U(\sim10$eV) prohibits the doubly occupied states.
Therefore the most important physics in our consideration
is the nearest neighbor repulsion between small polarons. 

In the lowest order approximation, we can neglect the DE term in the
above Hamiltonian, which then can be transformed into the spin one-half 
AFM Ising model,
\begin{equation}
{\cal H}_c=\frac{V}{4}\sum_{<ij>}\sigma_i\sigma_j,
\end{equation}
where we have used the pseudospin operator
($c_i^{\dagger}c_i=\frac{1}{2}+\frac{1}{2}\sigma_i$)\cite{Yamada}.
The AFM transition of the above Ising model should be
interpreted as the CO transition, and  the sublattice magnetization,
$\langle\sigma_i\rangle$ will correspond
to the order parameter of a given charge order-disorder transition.
The spontaneous magnetization will occur
below the transition temperature $T_N$.
In the mean field approximation, $T_N$ is given by 
$zV/4$ ($z$ : coordination number) which 
corresponds to the CO transition temperature $T_{CO}$.
Thus below $T_{CO}$, the real space ordering of Mn$^{3+}$/Mn$^{4+}$ species
takes place, minimizing the repulsion between localized small polarons.

Another notable thing in the original Hamiltonian, Eq.(1)
is that the CO state becomes unstable
when $V\lesssim t\langle {\rm cos}\frac{\theta}{2}\rangle$. 
Since both $V$ and $t\langle {\rm cos}\frac{\theta}{2}\rangle$ 
critically depend on the spin structure
and the electron-phonon interaction of the system, 
such a situation could happen 
by controlling external thermodynamic variables, {\it e.g.} magnetic field 
or pressure.  The subtle balance between two parameters, 
$V$ and $t\langle {\rm cos}\frac{\theta}{2}\rangle$,
will actually determine the stable ground state \cite{AHM}.  
The melting of the CO state induced by the magnetic field could be 
understood in this framework. This part will be left 
for future study.

To investigate lattice dynamics near the CO transition, 
we consider the extended Hamiltonian ${\cal H}_{cp}$, incorporating 
the electron-acoustic phonon coupling,
\begin{equation}
{\cal H}_{cp}= \frac{V}{4}\sum_{<ij>}\sigma_i\sigma_j
        +\frac{1}{2}\sum_{\vec{k}}(P_{\vec{k}}^{\ast}P_{\vec{k}}
         +\omega_0^2(k)Q_{\vec{k}}^{\ast}Q_{\vec{k}})
        +\frac{1}{\sqrt{N}}\sum_i\sum_{\vec{k}}\omega_0(k)g(k)
         \sigma_iQ_{\vec{k}}e^{i\vec{k}\cdot\vec{R}_i},
\end{equation}
where $P_{\vec{k}}$ and $Q_{\vec{k}}$ are the momentum and 
the amplitude of the local 
vibrational mode, and the final term represents the electron-phonon
interaction of local deformation type. 
We have adopted the pseudospin operator here too.
By employing the canonical transformation, 
$\bar{Q}_{\vec{k}}=Q_{\vec{k}}+\frac{g(k)}{\omega_0(k)}\sigma_{\vec{k}}$,
the electron-phonon coupling term can be decoupled.
Then introducing the generalized susceptibility function \cite{Yamada}, 
one can obtain the following renormalized phonon frequency,
$\tilde{\omega}(k)=\omega_0(k)/\sqrt{1+g^2(k)\chi_{\sigma\sigma}(\vec{k})}$,
where 
$\chi_{\sigma\sigma}(\vec{k})$
is an AFM spin susceptibility of the wave vector $\vec{k}$.

Since we are dealing with the acoustic phonon with $\omega_0(k)=sk$,
the renormalized sound velocity 
$\tilde{s}$ in the limit of $k\rightarrow 0$ is given by
\begin{equation}
\tilde{s}=s/\sqrt{1+g^2(0)\chi_{\sigma\sigma}(T)}.
\end{equation}
Here $\chi_{\sigma\sigma}(T)$ corresponds to the uniform AFM spin
susceptibility of the Ising model with the 
exchange interaction 
$\frac{1}{4}\tilde{V}=\frac{1}{4}\sum_{\vec{k}}(V_k-2g^2(k))
              e^{-i\vec{k}\cdot\vec{\tau}}$ 
($\vec{\tau}$ : nearest neighbor vector). 
Note that, as the electron-phonon coupling is turned on,
the AFM Neel temperature becomes reduced to 
$T_N=z\tilde{V}/4 (=T_{CO})$.
By using the uniform AFM spin susceptibility in the mean field 
approximation\cite{AFMsus}, the behavior of  sound velocity $\tilde{s}$ 
as a function of $T$ can be obtained (see Fig.~\ref{sv}). 
Due to a cusp maximum in the AFM susceptibility, there appears a
cusp minimum in the sound velocity at $T_{CO}$. The sound velocity 
is softened  smoothly above $T_{CO}$, but becomes substantially hardened 
below $T_{CO}$.  Quite interestingly, the behavior of
the above sound velocity around $T_{CO}$ is qualitatively very similar to 
the observation of Ramirez {\it et al.}\cite{Ramirez} 
(denoted by full dots in Fig.~\ref{sv}) for 
La$_{\rm 1-x}$Ca$_{\rm x}$MnO$_3$ ($0.63\leq{\rm x}\leq0.67$). 
It is noteworthy to compare the present sound velocity results 
with those in other charge ordered materials such as 
La$_{1.67}$Sr$_{0.33}$NiO$_4$ ($T_{CO}\sim 240$K)\cite{Ramirez2}.
The sound velocity just near $T_{CO}$ in this compound is
very similar to the present calculation, but 
as $T$ decreases far below $T_{CO}$,
it becomes again softened. That seems due to the residual charge hopping or
other possible effects in nickelates. On the contrary, 
in the manganese oxides, the residual hopping is not available
because the DE is suppressed by the CO.

The hardening of the sound velocity has been also reported
for the low doping compound (La$_{2/3}$Ca$_{1/3}$MnO$_3$)
of the FM metallic phase below
the FM transition temperature, $T_{C}$\cite{Ramirez,Jeong}.
But the hardenings of the sound velocity in two cases have different origins.
In the low doping case, the renormalization of the sound velocity occurs
due to $T$-dependent electron screening coming from the DE interaction, 
and so the hardening is enhanced by the external 
magnetic field which greatly influences the DE interaction\cite{Lee}. 
On the contrary, in the high doping case,
the present study reveals that the ordering of localized polarons 
is responsible for the renormalization.
Accordingly, the sound velocity in the CO state will not be susceptible to the 
external magnetic field, because the DE interaction is not involved in
the ordering and the accompanying renormalization. 
It is thus expected that the behavior of the sound velocity
under the magnetic field would be very different between two cases. 
Different behaviors for the above two cases were indeed observed in
the experiment of Ramirez {\it et al.}\cite{Ramirez}.
Meanwhile, the stronger magnetic field will destabilize the CO state
and definitely modify the behavior of the sound velocity
in the CO state, which needs further experimental investigations.

%
%


Now let's examine the magnetic structure in the CO state.
Collecting relevant spin interactions, we have 
\begin{equation}
{\cal H}_m=V\sum_{<ij>}n_in_j+t\sum_{<ij>}{\rm cos}\frac{\theta_{ij}}{2}
         c_i^{\dagger}c_j+J_s\sum_{<ij>}{\bf S}_i\cdot{\bf S}_j,
\end{equation}
where the last term is the AFM superexchange between local spins.
Although the DE interaction does not contribute to realizing the CO state,
it is the lowest order in the local spin interaction which yields
a direct connection between the spin and the CO ordering.
One can get the coupling between $e_g$ electrons 
and $t_{2g}$ local spins by expressing the DE,
$
{\cal H}_{DE} \simeq \frac{t}{\sqrt{2}}\sum_{<ij>}
              \left(1+\frac{{\bf S}_i\cdot{\bf S}_j}{2S^2}\right)
               c_i^{\dagger}c_j.
$
This approximation will be valid when $T$ is not too close to
$0$ K\cite{Millis}. Then by using the mean field
approximation; ${\cal H}_{DE} \simeq \frac{t}{2\sqrt{2}S^2}\sum_{<ij>}
              \langle c_i^{\dagger}c_j\rangle {\bf S}_i\cdot{\bf S}_j $, 
one can study effects of electron orderings 
on the local spins \cite{Bao}.

Defining 
$
J^{\prime}=\frac{t}{2\sqrt{2}S^2}\langle c_i^{\dagger}c_{i+\vec{\tau}}\rangle
$
$
(\vec{\tau}=\pm a\hat{x},\pm a\hat{y},\pm a\hat{z})
$
as the spin exchange coming from ${\cal H}_{DE}$,
we investigate how differently the exchange 
$J^{\prime}$ behaves between the disordered (uniform) phase and 
the ordered phase (CO).
The exchange $J_1^{\prime}$ in the disordered phase ($T>T_{CO}$)
corresponds to
$\frac{t}{2\sqrt{2}S^2}\frac{1}{N}
              \sum_{\vec{k}}\langle c_{\vec{k}}^{\dagger}c_{\vec{k}}\rangle
              e^{i\vec{k}\cdot\vec{\tau}}$ ($N$: number of sites),
which 
is nothing but the conventional FM interaction coming from the DE.
The situation, however, becomes quite distinct in the ordered 
phase ($T<T_{CO}$).
Below $T_{CO}$, the charge modulation starts to
occur as a result of the CO, and so
one can express $\langle c_i^{\dagger}c_{i+\vec{\tau}}\rangle$ as 
\begin{equation}
\langle c_i^{\dagger}c_{i+\vec{\tau}}\rangle=\frac{1}{N}
        \sum_{\vec{k}}\langle c_{\vec{k}}^{\dagger}c_{\vec{k}}\rangle
        e^{i\vec{k}\cdot\vec{\tau}} + \frac{1}{N}
        \sum_{\vec{Q}\in\Lambda}\sum_{\vec{k}}
        \langle c_{\vec{k}}^{\dagger}c_{\vec{k}+\vec{Q}}\rangle
        e^{i\vec{k}\cdot\vec{\tau}}
        e^{i\vec{Q}\cdot\vec{R}_i}e^{i\vec{Q}\cdot\vec{\tau}},
\end{equation}
where the second term results from the CO state, and
$\langle c_{\vec{k}}^{\dagger}c_{\vec{k}+\vec{Q}}\rangle$
is related to the CO order parameter with 
the charge modulation vector $\vec{Q}$.
If the carriers in the CO phase  are located in every other sites,
the vector $\vec{Q}$ belongs to the set of 
$\Lambda=\{\frac{\pi}{a}(\pm 1,\pm 1,\pm 1)\}$. 
As a result, the exchange $J^{\prime}$ is given by 
$
J^{\prime}=J_1^{\prime}+J_2^{\prime}
$
with
\begin{equation}
J_2^{\prime}=-\frac{t}{2\sqrt{2}S^2}\frac{1}{N}
             \left[\sum_{\vec{Q}\in\Lambda}\sum_{\vec{k}}
             \langle c_{\vec{k}}^{\dagger}c_{\vec{k}+\vec{Q}}\rangle
             e^{i\vec{k}\cdot\vec{\tau}}\right]e^{i\vec{Q}\cdot\vec{R}_i},
\end{equation}
for a given $\vec{\tau}$ ($e^{i\vec{Q}\cdot\vec{\tau}}=-1$) and
the lattice site vector $\vec{R}_i=a(l\hat{x}+m\hat{y}+n\hat{z})$.
Note that $J_2^{\prime}$ gives the AFM exchange interaction.
The strength of $J_2^{\prime}$ in the CO phase is expected to be comparable 
to those of $J_1^{\prime}$ and $J_s$, 
because $J_2^{\prime}$ is somehow related 
to the order parameter $\langle\sigma_i\rangle$ which 
increases rapidly below $T_{CO}$.
The phase factor 
$e^{i\vec{Q}\cdot\vec{R}_i}(=e^{i\pi(l+m+n)})$ in the spin exchange
$J_2^{\prime}$ of Eq.(7) is of particular interest, because
it gives rise to an unusual AFM spin structure.
Even though the explicit value of $J_2^{\prime}$ is hard to evaluate,
the schematic spin structure arising from $J_2^{\prime}$ can be obtained as
in Fig.~\ref{afm}(a) which shows the structure in $a$-$b$ plane. 
It is seen that the spin structure is
rather complicated with zig-zag shaped ferromagnetic chains which are 
mutually coupled in an antiferromagnetic way. 
Most remarkable is that this spin structure
is very similar to the real spin structure
of CE-type AFM (see Fig.~\ref{afm}(b)) observed in half-doped manganites
\cite{Wollan}. 
This comparison illustrates that the AFM exchange ($J_2^{\prime}$) 
induced by the CO 
may be essential in realizing the CE-type spin structure through
the competition with other exchanges such as the FM-DE ($J_1^{\prime}$), and
the already existing AFM superexchange ($J_s \sim$10 meV). 

The above arguments reflect that the DE interaction still plays a
essential role in the magnetism of the highly doped manganites,
and in this manner, one can understand consistently magnetic properties of
both metallic FM states ($0.2<{\rm x}<0.5$)
and the AFM ordered electron lattices (${\rm x}\gtrsim 0.5$)
within the same framework.
Finally it should be pointed out that the CO transition is likely to
be accompanied by the orbital ordering and the lattice distortion
\cite{Goodenough}. 
The latter phenomenon can be described in the present scheme by following the
procedure by Bari \cite{Bari}. The former, however, is not considered
in the present model, since the non-degenerate single band is
assumed for the DE interaction. 
Nonetheless, the orbital ordering may play an important role 
in determining the observed magnetic structure \cite{Goodenough,Kugel}. 
Generalized model including both the charge ordering and
the orbital ordering interaction
will certainly provide more realistic magnetic structure.

%
%


To summarize, we have described the CO phase in R$_{1/2}$A$_{1/2}$MnO$_3$ 
in the context of the order-disorder transition of
repulsively interacting small polarons.
We have found that the drastic renormalization of the sound velocity 
in the CO transition arises from the strong coupling of acoustic phonons
to the CO states. Further, we have found interesting implication of the
effect of the CO state on the magnetic structure, which may be essential 
for the observed specific CE-type AFM structure in the CO phase of half-doped
manganites.

Acknowledgments $-$
Helpful discussions with G. Khaliullin and A.M. Oles are greatly appreciated.
This work was supported by the POSTECH-BSRI program of the KME and
the POSTECH special fund, and in part by the KOSEF fund (K96176).
One of us (BIM) would like to thank L. Hedin and the MPI-FKF for
the hospitality during his stay.

\begin{figure}
\caption{The behavior of the renormalized sound velocity $\tilde{s}$
around the charge ordering (x=0.5).  In the inset, the AFM spin
susceptibility $\chi_{\sigma \sigma}(T)$ of the Ising model is plotted.
To help understanding, the comparison of present results
with Ramirez {\it et al.}'s (x = 0.63) (full dots) are also provided (Ref.[5]).
In this plot, we used $g^2(0)=0.8z\tilde{V}$ for the parameter of
the electron-phonon coupling strength.
}
\label{sv}
\end{figure}

\begin{figure}
\caption{(a) The AFM spin structure ($a$-$b$ plane) induced by the 
additional exchange $J_2^{\prime}$ in the CO phase. 
White and grey circles denote Mn$^{3+}$ and Mn$^{4+}$, respectively.
(b) The real spin ground state (CE-type) observed in the half-doped 
manganites (Ref.[3]).
In both cases, the spin 
structures are given by zig-zag shaped ferromagnetic chains
coupled antiferromagnetically.
}
\label{afm}
\end{figure}


\begin{references}
\bibitem{AHM} On certain conditions, Eq.(1) can be transformed 
into anisotropic Heisenberg Hamiltonian with spin one-half.
Then the transition between stable ground states can be described 
by the competition between two exchange parameters, the longitudinal 
$J_z$ and the transverse $J_{xy}$.
\end{references}
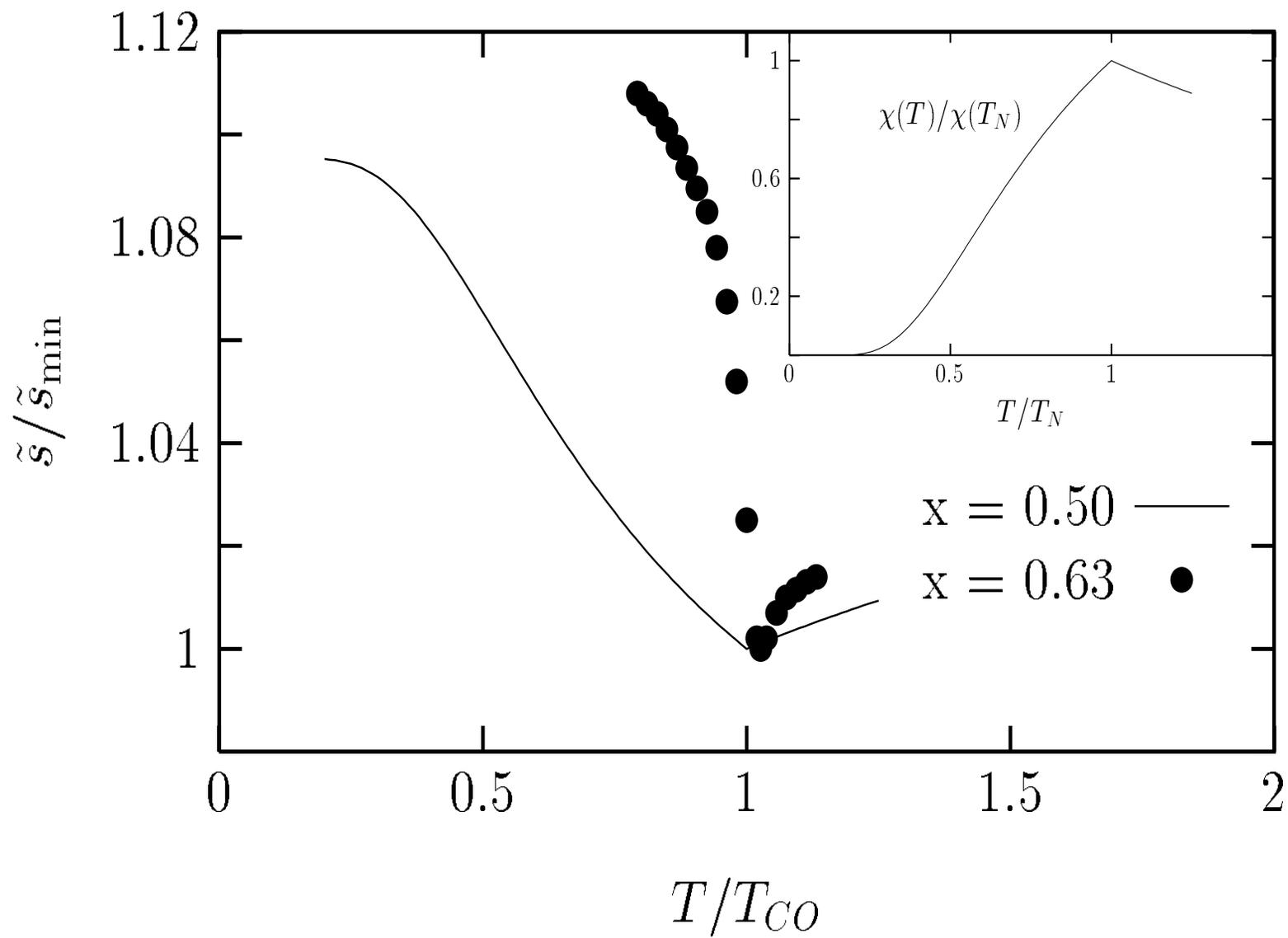
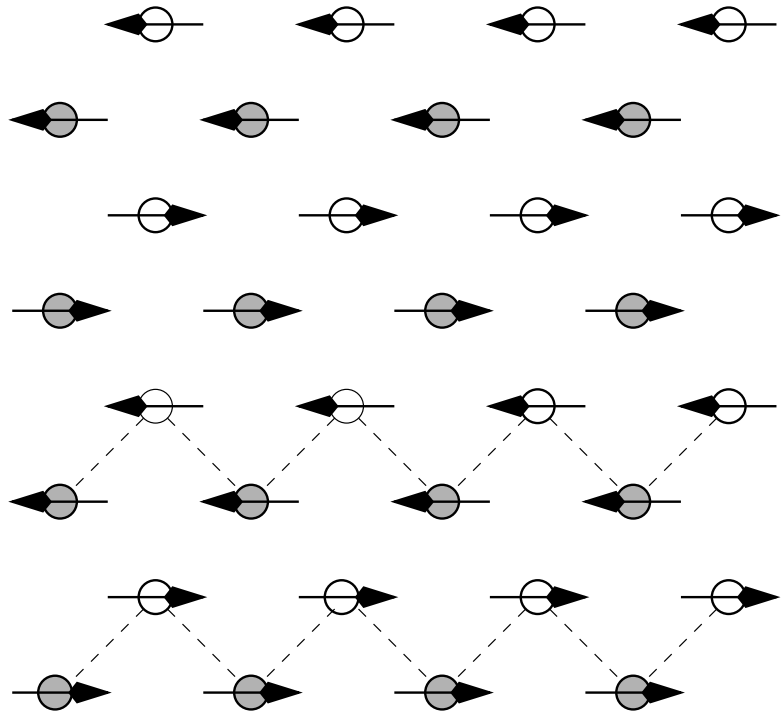
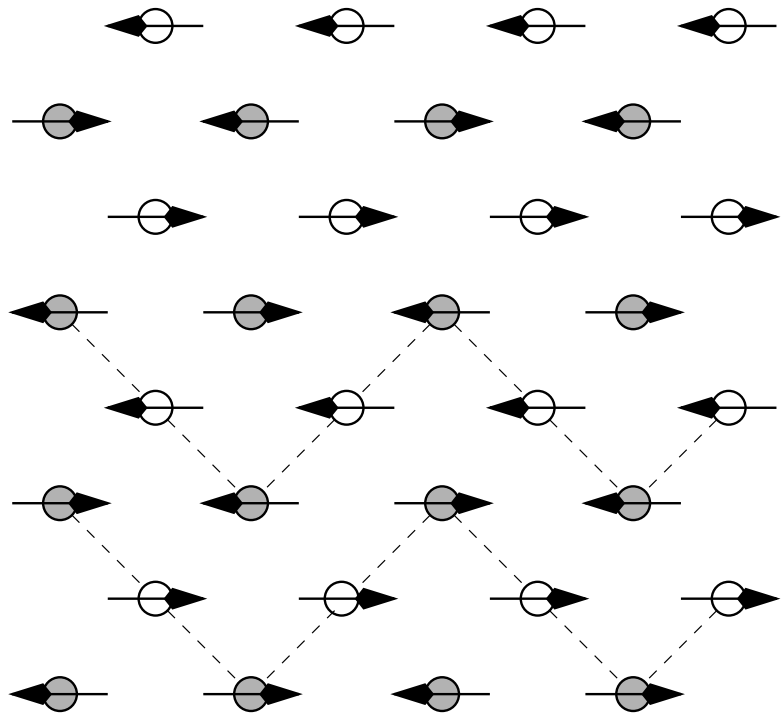
\end{document}